\documentclass[twocolumn,aps,prd,showpacs,preprintnumbers,amsmath,amssymb,showpacs,showkeys,nofootinbib]{revtex4}
\usepackage{latexsym,epsfig,amsmath,amssymb}
\usepackage{amssymb}
\usepackage{amsmath}
\usepackage{amsfonts}
\usepackage{epsfig}
\usepackage{verbatim}

\newcommand{\eq}{\begin{eqnarray}}
\newcommand{\en}{\end{eqnarray}}
\newcommand{\bea}{\begin{eqnarray}}
\newcommand{\eea}{\end{eqnarray}}

\newcommand{\ra}{\rangle}
\newcommand{\la}{\langle}

\begin{document}

\title{$B_{s(d)} - \bar B_{s(d)}$ mixing constraints on
flavor changing decays of $t$ and $b$ quarks}
\author{
Amand Faessler$^1$,
Thomas Gutsche$^1$,
Sergey Kovalenko$^2$,
Valery E. Lyubovitskij$^1$
\footnote{On leave of absence
from Department of Physics, Tomsk State University,
634050 Tomsk, Russia},
Ivan Schmidt$^2$
\vspace*{1.2\baselineskip}}
\affiliation{$^1$ Institut f\"ur Theoretische Physik,
Universit\"at T\"ubingen, \\
Kepler Center for Astro and Particle Physics,
\\ Auf der Morgenstelle 14, D-72076 T\"ubingen, Germany
\vspace*{1.2\baselineskip} \\
\hspace*{-1cm}$^2$ Departamento de F\'\i sica, Instituto de Estudios
Avanzados en Ciencias e Ingenier\'{\i}a y Centro Cient\'\i fico
Tecnol\'ogico de Valpara\'\i so (CCTVal), Universidad T\'ecnica
Federico Santa Mar\'\i a, Casilla 110-V, Valpara\'\i so, Chile
\vspace*{0.3\baselineskip}\\}

\date{\today}

\begin{abstract}

We study those dimension 6 effective operators which generate 
flavor-changing quark-gluon transitions of the third generation quarks,
with $t \rightarrow g +  u(c)$ and $b \rightarrow g + d(s)$, and 
which could be of interest for LHC experiments. We analyze the
contribution of these operators to $B_{s(d)} - \bar B_{s(d)}$ mixing
and derive limits on the corresponding effective couplings from the
existing experimental data. The Standard Model gauge invariance
relates these couplings to the couplings controlling $t\rightarrow g
+ u(c)$. On this basis we derive upper limits for the branching
ratios of these processes. We further show that forthcoming LHC
experiments might be able to probe the studied operators and the
physics beyond the Standard Model related to them.

\end{abstract}



\maketitle

The top quark is the least studied of the known quarks. Being the
heaviest it may offer new ways of probing physics beyond the
Standard Model (SM). The flavor-changing quark-gluon interactions
leading to the decays $t\rightarrow g + u(c)$ and $b\rightarrow g +
d(s)$ are examples for processes which are extremely suppressed in
the SM, and therefore experimental observation of  these decays
would be a smoking gun for new physics. In the following we study
these interactions within an effective Lagrangian approach. The most
general effective operators of the lowest dimension 6 representing
these interactions
are~\cite{Buchmuller:1985jz,AguilarSaavedra:2008zc}:
\eq
\label{operator-1}
\hspace*{-.4cm}
O_{qG}^{ij} &=& i \bar q_{iL} \lambda^A\,
\gamma^\mu D^\nu q_{jL} G^A_{\mu\nu} =  \\
\nonumber
&=& i\left(\bar u_{iL} \lambda^A \gamma^\mu D^\nu u_{jL} +
\bar d_{iL} \lambda^A \gamma^\mu D^\nu d_{jL} \right) G^A_{\mu\nu},  \\
\label{operator-2}
\hspace*{-.4cm}
O_{qG\phi}^{ij} &=& \bar q_{iL} \lambda^A
\sigma_{\mu\nu} d_{jR}  \phi  G^A_{\mu\nu} \rightarrow v\,
\bar d_{iL} \lambda^A \sigma_{\mu\nu} d_{jR} G^A_{\mu\nu},\\
\label{operator-3}
\hspace*{-.4cm}
O_{uG}^{ij} &=& i \bar u_{iR} \lambda^A\,
\gamma^\mu D^\nu u_{jR} G^A_{\mu\nu}, \\
\label{operator-4}
\hspace*{-.4cm}
O_{uG\phi}^{ij} &=& \bar q_{iL} \lambda^A
\sigma_{\mu\nu} u_{jR}  \tilde{\phi}  G^A_{\mu\nu} \rightarrow
v\, \bar u_{iL} \lambda^A \sigma_{\mu\nu} u_{jR} G^A_{\mu\nu}.
\en 
Here $G^A_{\mu\nu}$ and $\phi$ are the gluon and Higgs fields,
respectively; $\tilde\phi^j = \phi_i \epsilon^{ij}$, where
$\epsilon^{ij}$ is the antisymmetric tensor; $q_{iL}$ and $d_{iR}$
are the notations for the left-handed doublet and the right-handed
down quark of the $i$-th generation. The form of the operators to
the right of the arrows is taken after spontaneous symmetry
breaking. For the vacuum expectation of the Higgs field $\phi$ we
use $v = \la \phi \ra = 174$~GeV~\cite{Buchmuller:1985jz}. The
operator~(\ref{operator-1}) is the only one contributing both in the
up and down quark sectors. It generates interactions in both sectors
with the same coupling, as required by gauge invariance. Thus bounds
on the flavor violating coupling of the $b$-quark from low-energy
$B$-meson phenomenology would lead to the same constraints on the
corresponding couplings of the top quark.  The latter contributes to
the $t \to g + u(c)$ transition, which could be of interest for
LHC experiments. The operators~(\ref{operator-3})
and~(\ref{operator-4}) also contribute to these decays. 
However the low-energy constraints on their couplings could be deduced 
only at the loop-level~\cite{Loop-contr}. 
The second operator~(\ref{operator-2}), despite 
that it is not related to the top decays, could also be interesting
from the viewpoint of $B\rightarrow X_{s} g$ transitions at
$B$-factories~\cite{Drobnak:2010wh}.

In the present paper we derive constraints on the
operators~(\ref{operator-1}) and (\ref{operator-2}) from
the experimental data on the $B_{s(d)} - \bar B_{s(d)}$ mass
differences~\cite{Amsler:2008zz}-\cite{Abulencia:2006mq}:
\eq\label{Data}
& &\Delta m_{B_d} = 0.507 \pm 0.005  \ {\rm ps}^{-1}  \,, \nonumber\\[2mm]
& & 17 \ {\rm ps}^{-1} < \Delta m_{B_s} < 21 \ {\rm ps}^{-1} \,, \\[2mm]
& & \Delta m_{B_s} = 17.77^{+0.10}_{-0.10} \pm 0.07 \ {\rm ps}^{-1}
\, . \nonumber \en These data had a strong impact on the
phenomenology for physics beyond the SM (for a review see e.g.
Refs.~\cite{Lenz:2006hd}-\cite{Blanke:2008zb} and references
therein).

The $B_{s(d)} - \bar B_{s(d)}$ meson mass difference is related to
the matrix element of an effective Hamiltonian involving the $b \to \bar{b}$
transition~\cite{Okun:1982ap,Buchalla:1995vs}:
\eq\label{mass_diff}
\Delta m_{B_q} = 2 | \la \bar B_q | {\cal H}_{\rm eff} | B_q \ra | \,.
\en
The operators  (\ref{operator-1}) and (\ref{operator-2}) are constrained
by these data since they contribute to ${\cal H}_{\rm eff}$.
This contribution appears in second order of perturbation theory of
the interaction Lagrangian
\eq\label{L_new_physics}
{\cal L} &=& \frac{1}{\Lambda^2} \sum\limits_{i=1,2}
 \Biggl[
 \alpha_{3i} O_{qG}^{3i}      +  \alpha_{i3} O_{qG}^{i3} +
 \beta_{3i}  O_{dG\phi}^{3i}  +  \beta_{i3}  O_{dG\phi}^{i3} \biggr]
\nonumber\\
         &+& {\rm H.c.}
\en
with dimensionless couplings $\alpha_{ij}$, $\beta_{ij}$ and the
new physics scale $\Lambda$. The corresponding diagrams are shown 
in Fig.~1. Similar contributions to $K-\bar{K}$, $D-\bar{D}$ and $B-\bar{B}$ 
mixing were analyzed in the literature in relation to the SM dipenguin 
operators~\cite{Dipenguin,Petrov:1997ch}. However, to the best of our 
knowledge this analysis has not yet been extended beyond the SM. 
Our approach is similar to Ref.~\cite{Petrov:1997ch}. 
It is also based on the direct analysis 
of the diagrams similar to Fig.~1 implying the perturbative QCD regime. 
This  is justified by the fact that the scale of the momentum transfer 
through the gluon is set by the heavy quark mass $Q^{2}\sim -m_{b}^{2}$, 
which is large in comparison with $\Lambda_{QCD} \sim 200$ MeV.
\begin{figure}[htb]
\centering{\
\epsfig{figure=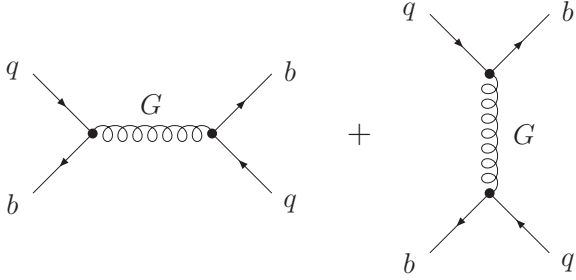,scale=.75}}
\vspace*{3cm}
\caption{Gluon-exchange diagrams contributing
to the mixing of neutral $B$ mesons.}
\label{fig:scattering}
\end{figure}
With the Lagrangian (\ref{L_new_physics}) we obtain (see Appendix) 
the \mbox{following} Hamiltonian terms
\eq\label{H_eff}
{\cal H}_{\rm eff} &=& {\cal H}_{\rm eff}^{(1)} \, + \,
{\cal H}_{\rm eff}^{(2)} \,, \\
{\cal H}_{\rm eff}^{(1)} &=&
  c_{31}^{11} Q_{11}^{bd} + c_{31}^{12} Q_{12}^{bd}
+ c_{32}^{11} Q_{11}^{bs} + c_{32}^{12} Q_{12}^{bs}
\, + \,  {\rm H.c.}\,, \nonumber\\
{\cal H}_{\rm eff}^{(2)} &=&
  c_{31}^{21} Q_{21}^{bd} + c_{31}^{22} Q_{22}^{bd} \nonumber\\
&+& c_{31}^{23} Q_{23}^{bd}
+ c_{32}^{21} Q_{21}^{bs} + c_{32}^{22} Q_{22}^{bs}
+ c_{32}^{23} Q_{23}^{bs}
\, + \,  {\rm H.c.}  \nonumber
\en
The effective couplings in Eq.~(\ref{H_eff}) are
expressed in terms of parameters of the underlying
Lagrangian (\ref{L_new_physics}):
\eq
\hspace*{-.3cm}
& &c_{3i}^{11} = c_{3i}^{12} = \frac{m_b^2}{\Lambda^4} \,
\alpha_{i3}^{\ast \, 2}\,, \\
\hspace*{-.3cm}
& &c_{3i}^{21} = - \frac{4 v^2}{\Lambda^4} \beta_{3i}^2\,,
   \hspace*{.1cm}
   c_{3i}^{22} = - \frac{4 v^2}{\Lambda^4} \beta_{i3}^{\ast \, 2}\,,
   \hspace*{.1cm}
   c_{3i}^{23} = - \frac{8 v^2}{\Lambda^4} \beta_{3i}
   \beta_{i3}^\ast \, . \nonumber
\en
The operators $Q_{ij}^{bq}$ can be expressed in terms of
operators of the so-called supersymmetric (SUSY)
basis~\cite{Gabbiani:1996hi}-\cite{Altmannshofer}:
\eq\label{SUSY_basis}
\hspace*{-.3cm}
& &O_1^{bq} = (\bar b_L^a \gamma^\mu q^a_L)
         (\bar b_L^b \gamma_\mu q^b_L)\,,
\hspace*{.3cm}
\tilde O_1^{bq} = (\bar b_R^a \gamma^\mu q^a_R)
         (\bar b_R^b \gamma_\mu q^b_R)\,, \nonumber\\
\hspace*{-.3cm}
& &O_2^{bq} = (\bar b_R^a q^a_L) (\bar b_R^b q^b_L)\,,
\hspace*{1cm}
\tilde O_2^{bq} = (\bar b_L^a q^a_R) (\bar b_L^b q^b_R)\,, \nonumber\\
\hspace*{-.3cm}
& &O_3^{bq} = (\bar b_R^a q^b_L) (\bar b_R^b q^a_L)\,,
\hspace*{1cm}
\tilde O_3^{bq} = (\bar b_L^a q^b_R) (\bar b_L^b q^a_R)\,, \nonumber\\
\hspace*{-.3cm}
& &O_4^{bq} = (\bar b_R^a q^a_L) (\bar b_L^b q^b_R)\,, \nonumber\\
\hspace*{-.3cm}
& &O_5^{bq} = (\bar b_R^a q^b_L) (\bar b_L^b q^a_A)\,,
\en
as
\eq\label{O_ij}
& &Q_{11}^{bq} = \frac{4}{3} O_1^{bq} \,, \nonumber\\
& &Q_{12}^{bq} = - \frac{2}{3} O_2^{bq} + 2 O_3^{bq} \,, \nonumber\\
& &Q_{21}^{bq} = \frac{4}{3} \tilde O_1^{bq}
               - \frac{2}{3} \tilde O_2^{bq}
               + 2 \tilde O_3^{bq} \,, \nonumber\\
& &Q_{22}^{bq} = \frac{4}{3} O_1^{bq}
               - \frac{2}{3} O_2^{bq}
               + 2 O_3^{bq} \,, \nonumber\\
& &Q_{23}^{bq} = - 4 \tilde O_2^{bq}
               + \frac{4}{3} \tilde O_3^{bq}
               - \frac{2}{3} O_4^{bq}
               + 2 O_5^{bq} \,,
\en
where $q=d$ or $s$; $a$ and $b$ are color indices.
The operators  $\tilde O_1^{bq}$ in Eq.~(\ref{SUSY_basis})
are obtained from $O_1^{bq}$ by exchanging $L \leftrightarrow R$.

For the calculation of the matrix elements of the effective Hamiltonian
${\cal H}_{\rm eff}$ we use the
relations~~\cite{Gabbiani:1996hi}-\cite{Becirevic:2001xt}:
\eq\label{matrix_elements}
\la \bar B_q | O_1^{bq}(\mu) | B_q \ra &=&
   \la \bar B_q | \tilde O_1^{bq}(\mu) | B_q \ra
=\frac{1}{3} m_{B_q} f_{B_q}^2 B_1^q(\mu) \,, \nonumber\\
\la \bar B_q | O_2^{bq}(\mu) | B_q \ra &=&
   \la \bar B_q | \tilde O_2^{bq}(\mu) | B_q \ra \nonumber\\
&=& - \frac{5}{24} \xi_{B_q}(\mu)  m_{B_q}
f_{B_q}^2 B_2^q(\mu) \,, \nonumber\\
\la \bar B_q | O_3^{bq}(\mu) | B_q \ra &=&
   \la \bar B_q | \tilde O_4^{bq}(\mu) | B_q \ra \nonumber\\
&=& \frac{1}{24} \xi_{B_q}(\mu)  m_{B_q}
f_{B_q}^2 B_3^q(\mu) \,, \\
\la \bar B_q | O_4^{bq}(\mu) | B_q \ra &=&
\frac{1}{4} \xi_{B_q}(\mu)  m_{B_q}
f_{B_q}^2 B_4^q(\mu) \,, \nonumber\\ \
\la \bar B_q | O_5^{bq}(\mu) | B_q \ra &=&
\frac{1}{12} \xi_{B_q}(\mu)  m_{B_q}
f_{B_q}^2 B_5^q(\mu) \,, \nonumber\\
\xi_{B_q}(\mu) &=& \biggl[\frac{m_{B_q}}{m_b(\mu)+m_q(\mu)}\biggr]^2
\,, \nonumber
\en
where $m_{B_q}$ and $f_{B_q}$ are the mass and decay constant 
of the $B_q$ meson. The $B_i(\mu)$ are the 
so-called ``bag''-parameters, which take into account the mismatch between
the vacuum saturation approximation (VSA) and the actual value for each
of the matrix elements (see detailed discussion in
Refs.~\cite{Gabbiani:1996hi}-\cite{Becirevic:2001xt}).
All $O_i^{bq}$ and $\tilde O_i^{bq}$ operators are renormalized
at the same scale $\mu = m_b$. Because of parity conservation in strong
interactions the matrix elements of the operators $\tilde O_i^{bq}$
and $O_i^{bq}$ coincide with each other~\cite{Ciuchini:1998ix}.

In the numerical calculations we use the following set of input parameters:
a renormalization scale parameter $\mu = m_b =~4.6$~GeV, quark masses
$m_d(\mu) = 5.4$~MeV, $m_s(\mu) = 150$~MeV,
$m_b(\mu) = 4.6$~GeV, $B$-meson masses and
decay constants
$m_{B_d} = 5.279$~GeV,
$m_{B_s} = 5.3675$~GeV,
$f_{B_d} = 189$~MeV,
$f_{B_s} = 230$~MeV. For the ``bag''-parameters
$B_i(\mu)$ we use the values computed in lattice QCD, with
Wilson fermions and with the nonperturbative regularization 
independent momentum subtraction (RI/MOM) renormalization 
scheme~\cite{Becirevic:2001xt}:
\eq\label{bag-parameters}
B_1^d(\mu) = 0.87, \hspace*{.3cm}
B_1^s(\mu) = 0.86 \,, \nonumber\\
B_2^d(\mu) = 0.82, \hspace*{.3cm}
B_2^s(\mu) = 0.83 \,, \nonumber\\
B_3^d(\mu) = 1.02, \hspace*{.3cm}
B_3^s(\mu) = 1.03 \,, \\
B_4^d(\mu) = 1.16, \hspace*{.3cm}
B_4^s(\mu) = 1.17 \,, \nonumber\\
B_5^d(\mu) = 1.91, \hspace*{.3cm}
B_5^s(\mu) = 1.94 \,. \nonumber
\en

Now we calculate the contribution of the effective Hamiltonian
${\cal H}_{\rm eff}$ of Eq. (\ref{H_eff}) to the mass difference.
The result is
\begin{eqnarray}\label{m_Bd}
\Delta m_{B_d} &=& \frac{2 |\alpha_{13}|^2}{9 \Lambda^4} \,
m_{B_d} \, m_b(\mu) \, f_{B_d}^2 \, B^d_{123}(\mu) \nonumber\\
&+& \frac{16 v^2}{9 \Lambda^4} \,
m_{B_d} \, f_{B_d}^2 \,
\biggl[ ( |\beta_{13}|^2 + |\beta_{31}|^2 ) \, B^d_{123}(\mu)
\nonumber\\
&+& 2 |\beta_{13}| |\beta_{31}| \, B^d_{45}(\mu) \biggr] \, ,\\
\label{m_Bs}
\Delta m_{B_s} &=& \frac{2 |\alpha_{23}|^2}{9 \Lambda^4} \,
m_{B_s}\, m_b(\mu) \, f_{B_s}^2 \, B^s_{123}(\mu) \nonumber\\
&+&\frac{16 v^2}{9 \Lambda^4} \,
m_{B_s} \, f_{B_d}^2 \,
\biggl[ ( |\beta_{23}|^2 + |\beta_{32}|^2 ) \, B^s_{123}(\mu)
\nonumber\\
&+& 2 |\beta_{23}| |\beta_{32}| \, B^s_{45}(\mu) \biggr] \, ,
\end{eqnarray}
where we introduced the following notations for the combinations of
``bag''-parameters:
\eq
B^q_{123}(\mu) &=& \biggl| 2 B_1^q(\mu) + \xi_{B_q}(\mu)
\biggl( \frac{5}{8}  B_2^q(\mu) + \frac{3}{8}  B_3^q(\mu) \biggr)
\biggr| \,, \nonumber\\
B^q_{45}(\mu) &=& \xi_{B_q}(\mu) \,
\biggl| - \frac{21}{4}  B_4^q(\mu) +
\frac{5}{4}  B_5^q(\mu) \biggr| \,.
\end{eqnarray}
>From Eqs.~(\ref{m_Bd}) and (\ref{m_Bs}) we derive the following
upper limits for the parameters of the  Lagrangian (\ref{L_new_physics}):
\eq
& &\frac{|\alpha_{13}|^2}{\Lambda^4} <
\frac{9 \, \Delta m_{B_d}}{2 \, m_b^2(\mu) \,
f_{B_d}^2 \, m_{B_d} \, B^d_{123}(\mu)}
\,,  \nonumber\\
& &\frac{|\alpha_{23}|^2}{\Lambda^4} <
\frac{9 \, \Delta m_{B_s}}{2 \, m_b^2(\mu) \,
f_{B_s}^2 \, m_{B_s} \, B^s_{123}(\mu)} \,, \\
& &\frac{|\beta_{13}|^2 \, v^2}{\Lambda^4} =
   \frac{|\beta_{31}|^2 \, v^2}{\Lambda^4} <
\frac{9 \, \Delta m_{B_d}}{16 \, f_{B_d}^2 \, m_{B_d} \, B^d_{123}(\mu)}
\,,  \nonumber\\
& &\frac{|\beta_{23}|^2 \, v^2}{\Lambda^4} =
   \frac{|\beta_{32}|^2 \, v^2}{\Lambda^4} <
\frac{9 \, \Delta m_{B_s}}{16 \, f_{B_s}^2 \, m_{B_s} \, B^s_{123}(\mu)}
\,. \nonumber
\en
Using data (\ref{Data}) we finally deduce the following bounds on the
coupling constants $\alpha_{ij}$ and $\beta_{ij}$:
\begin{eqnarray}\label{lim-1}
& &\frac{|\alpha_{13}|}{\Lambda^2} < 3.6 \times 10^{-7} \, \mbox{GeV}^{-2}\,,
\nonumber\\
& &\frac{|\alpha_{23}|}{\Lambda^2} < 1.7 \times 10^{-6}\, \mbox{GeV}^{-2}\,,
\label{lim-2} \\
& &\frac{|\beta_{13}| v}{\Lambda^2} =
   \frac{|\beta_{31}| v}{\Lambda^2} < 5.6 \times 10^{-7}\,  \mbox{GeV}^{-2}\,,
\nonumber\\
& &\frac{|\beta_{23}| v}{\Lambda^2} =
   \frac{|\beta_{32}| v}{\Lambda^2} < 2.8 \times 10^{-6}\,  \mbox{GeV}^{-2}\,.
\nonumber
\end{eqnarray}
The operator of Eq. (\ref{operator-1}) contains both
$b$ and $t$ quark terms with the same couplings to the gluon field, as 
dictated by the SM gauge invariance. Therefore, we can apply the above limits
to the derivation of the decay rate involving the top quark flavor-changing
neutral current (FCNC). The corresponding formula for the decay rate is:
\begin{eqnarray}\label{decay-rate-1}
\Gamma(t\rightarrow u_{i} g) = \frac{m_{t}^{5}}{12 \pi}
\frac{|\alpha_{3i} + \alpha^{*}_{i3}|^{2}}{\Lambda^{4}}
\end{eqnarray}
Using the limits of Eq. (\ref{lim-1}) we get for the branching ratios
\eq
\label{lim-3}
\frac{\Gamma(t\rightarrow u g)}{\Gamma_{t}} \leq 1.6 \times 10^{-3}, \ \
\frac{\Gamma(t\rightarrow c g)}{\Gamma_{t}} \leq 3.6 \times 10^{-2},
\en
where $\Gamma_{t}$ is the top quark total decay width which can be
approximated by the dominant mode~\cite{Chakraborty:2003iw}:
\eq
\Gamma_{t} \approx \Gamma(t\rightarrow b W) = 1.42 |V_{tb}|^{2} \,.
\en
These limits are to be compared to the existing
CDF~\cite{Incandela:1995jt} limit derived in~\cite{Han:1996ep}:
\eq\label{lim-4}
\frac{\Gamma(t\rightarrow c g)}{\Gamma_{t}}\leq 0.45.
\en
For the LHC experiments preliminary esti\-ma\-ti\-ons
gi\-ve~\cite{Ferreira:2005dr,Kim:2002de}:
\begin{eqnarray}\label{lim-5}
\frac{\Gamma(t\rightarrow u g)}{\Gamma_{t}} \leq 0.1,
\end{eqnarray}
which corresponds to a 10\% precision measurement of $\Gamma_{t}$.
As can be seen, this is not too far away from the limits of Eqs.~(\ref{lim-3}),
and with an improved precision on $\Gamma_{t}$ these FCNC transitions could
be probed by the LHC experiments. 

Note that above bounds~(\ref{lim-3}) are obtained with an ad hoc assumption 
about the vanishing contribution of the operators~(\ref{operator-3}) 
and~(\ref{operator-4}) to the decay rate~(\ref{decay-rate-1}).  
As we mentioned in the introduction they are not constrained by low-energy 
processes at tree-level. Therefore, taking  them into account may 
significantly relax the constraints~(\ref{lim-3}) essentially improving 
the prospects for searches involving the 
\mbox{$t\rightarrow u g$} transition.   

In conclusion, we analyzed a subset of effective dim=6 operators describing
flavor-changing interactions of the 3rd generation quarks with gluons,
representing one of the manifestations of physics beyond the SM.
We derived their contribution to the $B_{s(d)} - \bar B_{s(d)}$
mass difference and extracted upper limits
on the parameters of these operators from the experimental data.
With these limits we evaluated constraints on the branching ratios of
the top quark decays $t\rightarrow g + u(c)$, and found that the
LHC experiments have good prospects to probe the studied operators and the
new physics related to them.

\begin{acknowledgments}

This work was supported by the DFG under Contract No.FA67/31-2
and No.GRK683, by FONDECYT projects 1100582 and 110287, and
Centro-Cient\'\i fico-Tecnol\'{o}gico de Valpara\'\i so PBCT ACT-028.
This research is also part of the European
Community-Research Infrastructure Integrating Activity ``Study of
Strongly Interacting Matter'' (HadronPhysics2, Grant Agreement 
No.227431), Russian President grant ``Scientific Schools''
No.3400.2010.2, Russian Science and Innovations Federal Agency contract
No.02.740.11.0238.

\end{acknowledgments}

\appendix\section{Derivation of the second-order effective
Hamiltionian~${\cal H}_{\rm eff}$}

With the effective operators ~(\ref{operator-1}) and (\ref{operator-2})
we can generate at second-order of perturbation theory  matrix elements
describing the $s$- and $u$-channel quark transition
$q \, \bar b \, \to \, b \, \bar q$ (see diagrams in Fig.1)
\eq
M^{(1)} = M^{(1)}_s +  M^{(1)}_u,\ \ \ M^{(2)} = M^{(2)}_s +  M^{(2)}_u,
\en
where the first and the second matrix elements correspond to the
$O_{qG}$ and $O_{dG\phi}$ operators respectively.
Here the subscripts $s$ and $u$ denote the $s$- and $u$-channel contributions.
Below we show the results for the $s$-channel
contributions $M^{(1)}_s$ and $M^{(2)}_s$ [the crossing $u$-channel
results are obtained via the replacement $q_i \leftrightarrow - q_f$,
$u_q(q_i) \leftrightarrow v_q(q_f)$]:
\begin{widetext}
\eq\label{M1}
M^{(1)}_s &=& - \, \frac{\Gamma_{\mu\alpha}^{(1)}}{\Lambda^4 \, s} \,
\bar u_b(p_f) \lambda^A \gamma^\mu P_L v_q(q_f) \,
\bar v_b(p_i) \lambda^A \gamma^\alpha P_L u_q(q_i) \,, \nonumber\\
\Gamma_{\mu\alpha}^{(1)}  &=&
4 (\alpha_{3i} q_f^\nu - \alpha_{i3}^\ast p_f^\nu)
(\alpha_{3i} q_i^\beta - \alpha_{i3}^\ast p_i^\beta)
( g_{\nu\beta}  p_\mu p_\alpha + g_{\mu\alpha}  p_\nu p_\beta
- g_{\nu\alpha} p_\mu p_\beta - g_{\mu\beta}   p_\nu p_\alpha )
\en
and
\eq\label{M2}
M^{(2)}_s &=& M^{(2)}_{s, LL} +  M^{(2)}_{s, RR} +  M^{(2)}_{s, LR} +
M^{(2)}_{s, RL}  \,, \nonumber\\
M^{(2)}_{s, LL} &=&   \frac{v^2 \, \beta_{i3}^{\ast 2}}{\Lambda^4 \, s} \,
\Gamma_{\mu\nu;\alpha\beta}^{(2)} \
\bar u_b(p_f) \lambda^A \sigma^{\mu\nu} P_L v_q(q_f) \,
\bar v_b(p_i) \lambda^A \sigma^{\alpha\beta} P_L u_q(q_i) \,, \nonumber\\
M^{(2)}_{s, RR} &=&  \frac{v^2 \, \beta_{3i}^{2}}{\Lambda^4 \, s} \,
\Gamma_{\mu\nu;\alpha\beta}^{(2)}
\bar u_b(p_f) \lambda^A \sigma^{\mu\nu} P_R v_q(q_f) \,
\bar v_b(p_i) \lambda^A \sigma^{\alpha\beta} P_R u_q(q_i) \,, \nonumber\\
M^{(2)}_{s, LR} &=& \frac{v^2 \, \beta_{3i}\beta_{i3}^\ast}{\Lambda^4 \, s} \,
\Gamma_{\mu\nu;\alpha\beta}^{(2)}
\bar u_b(p_f) \lambda^A \sigma^{\mu\nu} P_L v_q(q_f) \,
\bar v_b(p_i) \lambda^A \sigma^{\alpha\beta} P_R u_q(q_i) \,, \\
M^{(2)}_{s, RL} &=&  \frac{v^2 \,
\beta_{3i}\beta_{i3}^\ast }{\Lambda^4 \, s}\,
\Gamma_{\mu\nu;\alpha\beta}^{(2)}
\bar u_b(p_f) \lambda^A \sigma^{\mu\nu} P_R v_q(q_f) \,
\bar v_b(p_i) \lambda^A \sigma^{\alpha\beta} P_L u_q(q_i) \,, \nonumber\\
\Gamma_{\mu\nu;\alpha\beta}^{(2)} &=&
4 ( g_{\nu\beta}  p_\mu p_\alpha + g_{\mu\alpha}  p_\nu p_\beta
- g_{\nu\alpha} p_\mu p_\beta - g_{\mu\beta}   p_\nu p_\alpha ) \,. \nonumber
\en
\end{widetext}
Here $P_L = (1 - \gamma_5)/2$, $P_R = (1 + \gamma_5)/2$;
$p_i(p_f)$ and $q_i(q_f)$ are the momenta of the bottom and the light quark
in the initial (final) state, respectively; 
$p$ is the intermediate gluon momentum.

For the derivation of the effective operators contributing to the
$B_{s(d)} - \bar B_{s(d)}$ mass difference we consider static 
limit for the $b$ quarks (their 3-momenta are equal to 
zero $\vec{p}_i = \vec{p}_f = 0$), which is well justified in the heavy 
quark limit $m_b \to \infty$. The momenta of quarks read as: 
$p_i = (m_b, \vec{0})\,,$ 
$p_f = (m_b, \vec{0})\,,$ 
$q_i = (E_i, \vec{q}_i)\,,$ and  
$q_f = (E_i, \vec{q}_f)\,,$   
where the energies and 3-momenta of the light quarks are of order of 
the constituent quark mass and are counted 
as ${\cal O}(1)$ in the heavy quark mass expansion. 
Then for the Mandelstam variables we get: 
\begin{eqnarray}\label{kinematics}
& & s = (p_i + q_i)^2 = (p_f + q_f)^2 = m_b^2 + m_q^2 + 2 m_b E_{q_i} \,, 
\nonumber\\
& & t = (p_i - p_f)^2 = (q_i - q_f)^2 = 0 \,, \nonumber\\
& & u = (p_i - q_f)^2 = (p_f - q_i)^2 = m_b^2 + m_q^2 - 2 m_b E_{q_i} \,, 
\nonumber\\
& & s + t + u = 2 (m_b^2 + m_q^2) \,.  
\end{eqnarray}
Note that the $u$-variable on general kinematical grounds can vanish 
at $E_{q} = (m_{b}^{2} + m_{q}^{2})/2m_{b}$, leading to the pole in the 
$u$-channel diagram and introducing an uncontrollable long-distance 
contribution. However, it is well known that the heavy quark carries 
nearly  the whole part of the heavy-light meson momentum, so that
the meson distribution amplitude, depending on the heavy quark momentum 
fraction $x$, is strongly peaked at $x\sim 1$~\cite{Petrov:1997ch}. 
The light quark momentum depends on the confinement potential and its 
typical average values lie around $0.5-0.6$~GeV or are even 
smaller~\cite{Voloshin}. Therefore, the relative contribution of the 
kinematical configuration leading to the $u$-pole is strongly suppressed 
and in a reasonable approximation we may neglect in Eqs.~(\ref{kinematics}) 
both the light quark mass $m_{q}$ and its energy $E_{q}$. A more accurate 
approach, based on pQCD, using model distribution amplitudes, was applied 
in Ref.~\cite{Petrov:1997ch} for the evaluation of the Standard Model 
dipenguin diagrams similar to those analyzed in the present paper. 
For our rough estimations we simply take: $s \simeq u \simeq m_b^2$.

Next, we simplify the
matrix elements (\ref{M1}) and (\ref{M2}) using  the equations of
motion for quark $u$ and antiquark $v$ spinors, applying the
heavy quark limit $m_q/m_b \ll 1$ and using the Fierz
identities for the spinor and color matrices:
\eq\label{Fierz_spin}
\hspace*{-.5cm}
& &(\gamma^\mu)_{\alpha\beta} \ (\gamma_\mu)_{\rho\sigma} \ = \
\delta_{\alpha\sigma} \ \delta_{\rho\beta}
 \ - \ (\gamma_5)_{\alpha\sigma} \ (\gamma_5)_{\rho\beta} -\nonumber\\
\hspace*{-.5cm} 
& &  \frac{1}{2}
 (\gamma^\mu)_{\alpha\sigma} \ (\gamma_\mu)_{\rho\beta}
 \ - \ \frac{1}{2}
 (\gamma^\mu\gamma_5)_{\alpha\sigma} \ (\gamma_\mu\gamma_5)_{\rho\beta}
\,, \\[3mm] 
\hspace*{-.5cm}
& &(\gamma^\mu \, P_{R/L})_{\alpha\beta} \
   (\gamma_\mu \, P_{R/L})_{\rho\sigma} = 
-  (\gamma^\mu \, P_{R/L})_{\alpha\sigma} \
   (\gamma_\mu \, P_{R/L})_{\rho\beta} \,, \nonumber\\[3mm] 
\hspace*{-.5cm}
& &(\gamma^\mu \, P_{R/L})_{\alpha\beta} \
   (\gamma_\mu \, P_{L/R})_{\rho\sigma} =
 2 \, (P_{L/R})_{\alpha\sigma} \
      (P_{R/L})_{\rho\beta} \, \nonumber
\en
and
\eq\label{Fierz_color}
\delta_{ab} \delta_{cd} &=&
\frac{1}{3} \delta_{ad} \delta_{cb} +
\frac{1}{2} \lambda_{ad}^A \lambda_{cb}^A  \,, \nonumber \\
\lambda_{ab}^A \lambda_{cd}^A &=&
\frac{16}{9} \delta_{ad} \delta_{cb} -
\frac{1}{3} \lambda_{ad}^A \lambda_{cb}^A  \,.
\en
With these identities we derive relations between the different four-quark
operators under investigation and express them in terms of the SUSY basis:
\eq
\bar b_{L}^a \gamma^\mu \lambda^A_{ab} q_{L}^b
\ \bar b_{L}^c \gamma_\mu \lambda^A_{cd} q_{L}^d
&=& \frac{4}{3} \bar b_{L}^a \gamma^\mu q_{L}^a
\ \bar b_{L}^b \gamma_\mu q_{L}^b \ = \ \frac{4}{3} O_1^{bq}
\,, \nonumber\\
\bar b_{R}^a \gamma^\mu \lambda^A_{ab} q_{R}^b
\ \bar b_{R}^c \gamma_\mu \lambda^A_{cd} q_{R}^d
&=& \frac{4}{3} \bar b_{R}^a \gamma^\mu  q_{R}^b
\ \bar b_{R}^b \gamma_\mu q_{R}^a  \ = \
\frac{4}{3} \tilde O_1^{bq}  \,, \nonumber\\
\bar b_{R}^a \gamma^\mu \lambda^A_{ab} q_{R}^b
\ \bar b_{L}^c \gamma_\mu \lambda^A_{cd} q_{L}^d
&=& - 2  \bar b_{R}^a \lambda^A_{ab} q_{L}^d
\ \bar b_{L}^c \lambda^A_{cd} q_{R}^b \nonumber\\
\label{Fin}
&=&
\frac{4}{3} O_5^{bq} - 4 O_4^{bq}   \,, \\
\bar b_{R}^a \lambda^A_{ab} q_{L}^b
\ \bar b_{R}^c \lambda^A_{cd} q_{L}^d
&=&- \frac{2}{3} O_2^{bq} + 2 O_3^{bq}  \,, \nonumber\\
\bar b_{L}^a \lambda^A_{ab} q_{R}^b
\ \bar b_{L}^c \lambda^A_{cd} q_{R}^d
&=&- \frac{2}{3} \tilde O_2^{bq} + 2 \tilde O_3^{bq}
\,, \nonumber\\
\bar b_{R}^a \lambda^A_{ab} q_{L}^b
\ \bar b_{L}^c \lambda^A_{cd} q_{R}^d
&=& - \frac{2}{3} O_4^{bq}  + 2  O_5^{bq}  \,. \nonumber
\en
Note that the contributions of the $s$- and $u$-channel
diagrams are equal to each other
within the approximations used in our analysis.
Finally, we derive the expressions for the effective Hamiltonians
${\cal H}_{\rm eff}^{(1)}$ and ${\cal H}_{\rm eff}^{(2)}$  corresponding
to the matrix elements $M^{(1)}$ and $M^{(2)}$.
The result is shown in  Eqs.~(\ref{H_eff})~-~(\ref{O_ij}).


\begin{thebibliography}{99}
\bibitem{Buchmuller:1985jz}
  W.~Buchmuller and D.~Wyler,
  Nucl.\ Phys.\ B {\bf 268}, 621 (1986).

\bibitem{AguilarSaavedra:2008zc}
  J.~A.~Aguilar-Saavedra,
  Nucl.\ Phys.\  B {\bf 812}, 181 (2009)
  [arXiv:0811.3842 [hep-ph]].
  
\bibitem{Loop-contr} A.~ Faessler, T.~Gutsche, J.~C.~Helo,
 S.~Kovalenko, V.~E.~Lyubovitskij, I.~Schmidt, in preparation.


\bibitem{Drobnak:2010wh}
  J.~Drobnak, S.~Fajfer and J.~F.~Kamenik,
  Phys.\ Rev.\ Lett.\  {\bf 104}, 252001 (2010)
  [arXiv:1004.0620 [hep-ph]].


\bibitem{Amsler:2008zz}
  C.~Amsler {\it et al.}  [Particle Data Group],
  Phys.\ Lett.\  B {\bf 667}, 1 (2008).
\bibitem{Abazov:2006dm}
  V.~M.~Abazov {\it et al.}  [D0 Collaboration],
  Phys.\ Rev.\ Lett.\  {\bf 97}, 021802 (2006)
  [arXiv:hep-ex/0603029].
\bibitem{Abulencia:2006mq}
  A.~Abulencia {\it et al.}  [CDF - Run II Collaboration],
  Phys.\ Rev.\ Lett.\  {\bf 97}, 062003 (2006); 
  [AIP Conf.\ Proc.\  {\bf 870}, 116 (2006)]
  [arXiv:hep-ex/0606027]; 
  M.~Jones  [CDF - Run II Collaboration],
  AIP Conf.\ Proc.\  {\bf 870}, 116 (2006).
\bibitem{Lenz:2006hd}
  A.~Lenz and U.~Nierste,
  JHEP {\bf 0706}, 072 (2007) \\{}
  [arXiv:hep-ph/0612167].
\bibitem{Ko:2008xb}
  P.~Ko and J.~H.~Park,
  Phys.\ Rev.\  D {\bf 80}, 035019 (2009)
  [arXiv:0809.0705 [hep-ph]].
\bibitem{Blanke:2008zb}
  M.~Blanke, A.~J.~Buras, B.~Duling, S.~Gori and A.~Weiler,
  JHEP {\bf 0903}, 001 (2009)
  [arXiv:0809.1073 [hep-ph]].
\bibitem{Okun:1982ap}
  L.~B.~Okun,
{\it ``Leptons And Quarks''},
Amsterdam, Netherlands: North-holland (1982) 361p.
\bibitem{Buchalla:1995vs}
  G.~Buchalla, A.~J.~Buras and M.~E.~Lautenbacher,
  Rev.\ Mod.\ Phys.\  {\bf 68}, 1125 (1996)
  [arXiv:hep-ph/9512380].
 
 \bibitem{Dipenguin} 
  J.~O.~Eeg and I.~Picek,
  Z.\ Phys.\  C {\bf 39}, 521 (1988); 
  Nucl.\ Phys.\  B {\bf 292}, 745 (1987);
Phys. Lett. B {\bf 160}, 154 (1985).

\bibitem{Petrov:1997ch}
  A.~A.~Petrov,
  Phys.\ Rev.\  D {\bf 56}, 1685 (1997)
  [arXiv:hep-ph/9703335].

\bibitem{Gabbiani:1996hi}
  F.~Gabbiani, E.~Gabrielli, A.~Masiero and L.~Silvestrini,
  Nucl.\ Phys.\ B {\bf 477}, 321 (1996)
  [arXiv:hep-ph/9604387].
\bibitem{Allton:1998sm}
  C.~R.~Allton {\it et al.},
  Phys.\ Lett.\ B {\bf 453}, 30 (1999)
  [arXiv:hep-lat/9806016].
\bibitem{Ciuchini:1998ix}
  M.~Ciuchini {\it et al.},
  JHEP {\bf 9810}, 008 (1998)
  [arXiv:hep-ph/9808328].
\bibitem{Becirevic:2001xt}
  D.~Becirevic, V.~Gimenez, G.~Martinelli, M.~Papinutto and J.~Reyes,
  JHEP {\bf 0204}, 025 (2002)
  [arXiv:hep-lat/0110091].
 \bibitem{Altmannshofer} W.~Altmannshofer, A.~J.~Buras, S.~Gori, 
  P.~Paradisi and D.~M.~Straub,
  Nucl.\ Phys.\  B {\bf 830}, 17 (2010) [arXiv:0909.1333 [hep-ph]].
\bibitem{Chakraborty:2003iw}
  D.~Chakraborty, J.~Konigsberg and D.~L.~Rainwater,
  Ann.\ Rev.\ Nucl.\ Part.\ Sci.\  {\bf 53}, 301 (2003)
  [arXiv:hep-ph/0303092];
  W.~Wagner,
  Rept.\ Prog.\ Phys.\  {\bf 68}, 2409 \\{} (2005)
  [arXiv:hep-ph/0507207];
  A.~Denner and T.~Sack, \\{} 
  Nucl.\ Phys.\  B {\bf 358}, 46 (1991);
  G.~Eilam, R.~R.~Mendel, R.~Migneron and A.~Soni, 
  Phys.\ Rev.\ Lett.\  {\bf 66}, 3105 \\{} (1991); 
  A.~Czarnecki and K.~Melnikov,
  Nucl.\ Phys.\  B {\bf 544}, 520 (1999)
  [arXiv:hep-ph/9806244]; \\{} 
  K.~G.~Chetyrkin, R.~Harlander, T.~Seidensticker and M.~Steinhauser,
  Phys.\ Rev.\  D {\bf 60}, 114015 (1999) \\{} 
  [arXiv:hep-ph/9906273].
  
  
\bibitem{Incandela:1995jt}
  J.~Incandela (CDF Collaboration),
  Nuovo Cim.\  {\bf 109A}, 741 (1996).

\bibitem{Han:1996ep}
  T.~Han, K.~Whisnant, B.~L.~Young and X.~Zhang,
  Phys.\ Rev.\  D {\bf 55}, 7241 (1997)
  [arXiv:hep-ph/9603247].

\bibitem{Ferreira:2005dr}
  P.~M.~Ferreira, O.~Oliveira and R.~Santos,
  Phys.\ Rev.\ D {\bf 73}, 034011 (2006)
  [arXiv:hep-ph/0510087].



\bibitem{Kim:2002de}
  Y.~K.~Kim  (CDF and D0 Collaborations),
  Int.\ J.\ Mod.\ Phys.\  A {\bf 17}, 3099 (2002).

\bibitem{Voloshin} M.~B.~Voloshin, Surveys \ in \ 
  High \ Energy \ Physics {\bf 8}, 27 (1995); 
  D.~S.~Hwang, C.~S.~Kim and W.~Namgung,
  Phys.\ Rev.\  D {\bf 54}, 5620 (1996)
  [arXiv:hep-ph/9604225]; 
  F.~De Fazio,
  Mod.\ Phys.\ Lett.\  A {\bf 11}, 2693 (1996)
  [arXiv:hep-ph/9608406].

\end{thebibliography}
\end{document}